\documentclass[conference,10pt]{IEEEtran}
\IEEEoverridecommandlockouts

\usepackage{array}
\usepackage{stfloats}
\usepackage{url}
\usepackage{verbatim}
\usepackage{algorithm}
\usepackage{subcaption}
\usepackage{hyperref}
\usepackage{cite}
\usepackage{amsmath,amssymb,amsfonts}
\usepackage{algorithmic}
\usepackage{graphicx}
\usepackage{textcomp}
\usepackage{xcolor}
\usepackage{comment}
\usepackage{caption} 
\usepackage{float}
\usepackage{multirow}
\usepackage{makecell}
\usepackage{enumitem}
\usepackage{bbm}

\def\SNR{\mathop{\mathrm{SNR}}} 

\begin{document}

\title{Optimization for Semantic-Aware Resource Allocation under CPT-based Utilities
}

\author{\IEEEauthorblockN{Symeon Vaidanis$^*$,  Photios A. Stavrou$^*$, and Marios Kountouris${^*}^{\dagger}$}\\
$^*$Communication Systems Department, EURECOM, Sophia-Antipolis, France\\
$^{\dagger}$Department of Computer Science and
Artificial Intelligence, University of Granada, Spain\\
Emails: \texttt{\{symeon.vaidanis, fotios.stavrou, marios.kountouris\}@eurecom.fr}}

\maketitle

\begin{abstract}
The problem of resource allocation in goal-oriented semantic communication with semantic-aware utilities and subjective risk perception is studied here. 
By linking information importance to risk aversion, we model agent behavior using Cumulative Prospect Theory (CPT), which incorporates risk-sensitive utility functions and nonlinear transformations of distributions, reflecting subjective perceptions of gains and losses. 
The objective is to maximize the aggregate utility across multiple CPT-modeled agents, which leads to a nonconvex, nonsmooth optimization problem. To efficiently solve this challenging problem, we propose a new algorithmic framework that combines successive convex approximation (SCA) with the projected subgradient method and Lagrangian relaxation,
Our approach enables tractable optimization while preserving solution quality, offering both theoretical rigor and practical effectiveness in semantics-aware resource allocation.
\end{abstract}


\section{Introduction}
Goal-oriented semantic communication has emerged as a transformative paradigm for future network design, focusing the entire communication process on the relevance and importance of information content, defined by application requirements and perceived by end users \cite{kountouris2021semantics,strinati:2024,stavrou:2023}. This framework minimizes redundant data transmission, enhances resource and computational efficiency, and serves as a cornerstone for enabling collaborative, hyperconnected intelligence and the emerging Internet of Agents.

A key challenge in semantic communication is to establish an operational and meaningful definition of information significance. In this context, messages are often subjectively evaluated and events interpreted on the basis of conceptual relevance and the decision maker's background knowledge.

To address this, we link the semantic importance of content with user behavior under risk, irrational biases in decision-making, and their subjective perception of event occurrence probabilities. To this end, we leverage concepts from cumulative prospect theory (CPT) \cite{Kahneman_Tversky_1992}, highlighting its potential for application in multi-objective optimization and semantics-aware resource allocation problems.

One of the main challenges in applying CPT lies in optimizing its utility function, which is inherently nonconvex and nonsmooth, even in its baseline formulation. 

Recent studies have primarily addressed this challenge in the context of portfolio optimization. In \cite{CPT_Portfolio_Boyd}, the authors employ a minorization–maximization (MM) algorithm to maximize the CPT utility, showing that the problem can be approached using a convex–concave procedure that iteratively optimizes a local approximation. However, the proposed methods are heuristic and lack formal theoretical guarantees.
In \cite{CPT_Portfolio_ADMM}, the authors adopt the Alternating Direction Method of Multipliers (ADMM) in two distinct approaches: one combined with dynamic programming, and the other with the Pool Adjacent Violators (PAV) algorithm. While effective in practice, these approaches, particularly the PAV-based method, fail to fully account for the inherent nonsmoothness and nonconvexity of the CPT utility function.

In this work, we propose an optimization framework for solving resource allocation problems in semantic communication involving CPT-based agents. We specifically apply our method to power allocation scenarios, as considered in \cite{ICC_2025_vaidanis}. The core contribution lies in explicitly addressing the non-smoothness and non-convexity of the utility function during the optimization process. To this end, we integrate the Successive Convex Approximation (SCA) \cite{SCA_Palomar,SCA_Scutari_2018,SCA_Scutari_Theory_Part1,SCA_Scutari_Applications_Part2}, the Projected Subgradient Method, and the Lagrangian relaxation \cite{boyd-book}. Numerical results demonstrate that the proposed method outperforms Sequential Quadratic Programming (as implemented in MATLAB's Optimization Toolbox) in terms of scalability with the number of agents, while incurring a slightly higher, yet comparable, execution time.

\section{CPT Preliminaries}
In this section, we provide a brief overview of the mathematical framework of CPT \cite{Wakker_Book}, introducing its key features absent in Expected Utility Theory (EUT).

First, agents evaluate events relative to a \textit{reference point} $x_0$, which may represent an expected or previously achieved operating level and can vary across different application scenarios. Deviations from $x_0$ reflect the agent’s subjective perception of value and form the basis for expressing preferences and decision biases, thus capturing the semantic relevance of the outcomes.

Second, the CPT utility function $u(x)$ exhibits \textit{loss aversion}, meaning that agents are more sensitive to losses than to equivalent gains. Formally, this is captured by the condition $u'(x_0^+) < u'(x_0^-)$, indicating a steeper slope on the loss side of the $x_0$.
\emph{Symmetric bet aversion} \cite{Kahneman_Tversky_1979} is defined by $u(x_0 + \delta) + u(x_0 -\delta) < 0, \forall \delta>0$, with $u(x_0)=0$. This condition implies that agents reject all fair symmetric gambles in favor of the status quo. A stronger form of this aversion requires $u(x_0 + \delta_1) + u(x_0 - \delta_1) < u(x_0 + \delta_2) + u(x_0 - \delta_2), \forall 0 < \delta_2 \leq \delta_1$, indicating a growing reluctance toward larger symmetric risks.
The curvature of $u(x)$ reflects both risk attitudes and sensitivity to change within each domain. In the gain domain, a concave function implies greater sensitivity to small gains near the reference point, indicating risk aversion, while a convex function suggests increased responsiveness to larger, distant gains, implying risk seeking. The opposite holds in the loss domain. This contrasts with EUT, which typically assumes a globally concave utility function. 

Third, CPT introduces \textit{nonlinear probability distortion}, where objective probabilities are transformed through a probability weighting function (PWF) to reflect subjective perception. The PWF captures how agents tend to overweight low-probability events and underweight moderate to high probabilities, modeling human biases in uncertainty assessment more accurately than traditional linear probability interpretations.

In this work, we adopt a recently proposed generalized formulation of CPT utility functions \cite{ICC_2025_vaidanis}, which allows the modeling of a wider spectrum of risk attitudes and behavioral patterns.
\begin{equation}
    u(x) = \left\{
        \begin{array}{lll}
            \lambda_1 \frac{\mu_1 - \exp\left( \frac{\alpha}{\gamma_1} \cdot \frac{(x - x_0)}{m}  \right)}{\alpha} & x_0 \leq x \\
            \lambda_2 \frac{\mu_2 - \exp\left( \frac{\beta}{\gamma_2} \cdot \frac{(x - x_0)}{n}  \right)}{\beta} & x < x_0 \\
        \end{array} 
    \right.
    \label{eq:KW_utility_extension}
\end{equation}
where $\alpha$, $\beta$, $\lambda_1$, $\lambda_2$, $\gamma_1$, $\gamma_2$, $\mu_1$, $\mu_2$, $m$ and $n$ are user specific parameters generally defined on $\mathbb{R}$. 

\begin{table}[h]
    \centering
    \begin{tabular}{|c|c|c|}
        \hline
         & Gain & Loss\\
        \hline
        Constant & \makecell{$\gamma_1 \to 0^-$,$0 < \alpha$, \\ $0 < \lambda_1 \cdot \mu_1$, $0<m$} & \makecell{$\gamma_2 \to 0^+$,$0 < \beta$, \\ $\lambda_2 \cdot \mu_2 < 0$,$0<n$} \\
        \hline
        Linear & $\alpha \to 0$,$\frac{\lambda_1}{\gamma_1}<0$, $0<m$ & $\beta \to 0$,$\frac{\lambda_2}{\gamma_2}<0$,$0<n$ \\
        \hline
        Convex & \makecell{$\frac{\lambda_1}{\gamma_1}<0$,$0<\frac{\alpha}{\gamma_1}$, \\ $\mu_1 \leq 1$, $0<m$} & \makecell{$\frac{\lambda_2}{\gamma_2}<0$,$0<\frac{\beta}{\gamma_2}$, \\ $1 \leq \mu_2$,$0<n$} \\
        \hline
        Concave & \makecell{$\frac{\lambda_1}{\gamma_1}<0$,$\frac{\alpha}{\gamma_1}<0$,\\ $1 \leq \mu_1$, $0<m$} & \makecell{$\frac{\lambda_2}{\gamma_2}<0$,$\frac{\beta}{\gamma_2}<0$,\\ $\mu_2 \leq 1$,$0<n$} \\
        \hline
    \end{tabular}
    \caption{Summary of parameter values and possible utility function shapes across subdomains.}
    \label{tab:parameters_utility_function}
\end{table}

\section{System Model and Problem Formulation}
Consider a system in which a server communicates with a set of agents $\mathcal{N} = \{1, \ldots, N\}$ over a wireless medium. Each agent exchanges information with the server, r, with varying degrees of semantic importance depending on its specific goal. We study a problem where resources have to be allocated among different agents based on the perceived importance of information, as evaluated using metrics derived from CPT.

For clarity of exposition, we consider the total transmit power $P_{total}$ as the resource to be allocated and model the CPT-based metric as a function of the communication channel conditions. These conditions are characterized by the signal-to-noise ratio (SNR), given by $\textrm{SNR} = \frac{P|h|^2}{\sigma^2}$, where $h$ denotes the channel coefficient, $P$ is the transmit power and $\sigma^2$ is the variance of the additive white Gaussian noise (AWGN). 

Each agent employs its own parametrized utility function, and the evaluation of the CPT-based metric under a given resource allocation is independent across agents, rendering the optimization problem separable. The total resource budget is finite, imposing an upper bound on the sum of allocated resources. In the most general case, the utility of each agent is weighted by a subjectively perceived probability, e.g., the likelihood that the $i$-th agent is active. As a result, the optimization problem is separable, but inherently non-concave and non-smooth, and can be formulated as follows:
\begin{equation}
    \begin{aligned}
        \min_{\mathbf{P}} \quad & f \left(\mathbf{P}\right)\\
        \textrm{s.t.} \quad & \mathbf{P} \in \mathcal{S}\\
          &  g \left(\mathbf{P}\right) \leq 0  \\
    \end{aligned}
    \label{eq:Opt_Problem_Generic}
\end{equation}
where $f \left(\mathbf{P}\right) = - \sum_{i=1}^{N}{w(p_i)f_i(P(i))} \;,\; f_i(P(i)) = u_i(\SNR(i)) \;,\; \SNR(i) = \frac{P(i) \cdot |h(i)|^2}{\sigma^2}$,$w(p_i)$ is the PWF, modeling the $i$-th agent's subjective assessment of probability $p_i$, $\mathcal{S} = \mathbb{R}^N_+$ and $g \left(\mathbf{P}\right) = \sum_{i=1}^{N}{P(i)} - P_{total}$.

\section{Optimization Problem}
In this section, we present a framework for solving general resource allocation problems that involve CPT-based agents. Our approach combines SCA with the PSM. We first present a general solution to the CPT-based resource allocation problem, which can then be particularized to specific scenarios, such as the power allocation problem. 
In that case, the parameters of the utility functions can take the following values: $\alpha(i)$, $\beta(i)  \in \mathbb{R}$ and $\lambda_1(i)$, $\lambda_2(i) \in \mathbb{R}^*$ and $\gamma_1(i)$, $\gamma_2(i) \in \mathbb{R}^*$ and $\mu_1(i) = \mu_2(i) = 1$ for $i \in \{1, \dots, N\}$. The reference point of the $i$-th agent is denoted as $\textrm{SNR}_0(i)$.

\subsection{Successive Convex Approximation}
The optimization problem in \ref{eq:Opt_Problem_Generic} can contain convex (from a maximization perspective) or linear components, depending on the values of the CPT parameters.
To handle this, we employ the SCA algorithm, which iteratively solves the original problem using surrogate functions. The SCA algorithm, adapted to the notation of the optimization problem \ref{eq:Opt_Problem_Generic}, is detailed in Algorithm \ref{alg:SCA}.
In essence, the SCA method approximates the concave (or convex) and linear parts of the objective function, with respect to whether the goal is maximization or minimization, by surrogate functions. The optimization problem is then solved iteratively, leveraging the results of the previous iteration at each step, until convergence is achieved.

The main construction rules for the surrogate function in the context of a minimization problem are as follows:
\begin{enumerate}
    \item $\Tilde{f}(\mathbf{x} | \mathbf{x}^{(k)})$ should be strongly convex over the domain $\mathcal{X}$.
    \item $\nabla \Tilde{f}(\mathbf{x}^{(k)} | \mathbf{x}^{(k)}) = \nabla f(\mathbf{x}^{(k)}), \; \forall \; \mathbf{x}^{(k)} \in \mathcal{X}$
    \item $\nabla\Tilde{f}(\mathbf{x} | \mathbf{x}^{(k)})$ must be continuous on $\mathcal{X}$
    \item \{Optional for stronger convergence guarantees\} $\nabla \Tilde{f}(\mathbf{x} | \mathbf{x}^{(k)})$ is uniformly Lipschitz continuous over $\mathcal{X}$
\end{enumerate}
where $\mathcal{X}$ denotes the domain of $\mathbf{x}$ and $\Tilde{f}(\mathbf{x} | \mathbf{x}^{(k)})$ is the surrogate function that approximates the original function $f(\mathbf{x})$ in the $(k+1)$-th iteration. It should be emphasized that these construction rules were originally proposed in \cite{SCA_Scutari_Theory_Part1} for smooth but nonconvex objective functions. However, the authors argue that these results can be extended to handle nonsmooth objectives as well. Specifically, if the objective function can be decomposed as the sum of a smooth but nonconvex function $U(\mathbf{x})$ (or concave, in the case of maximization) and a nonsmooth but convex function $R(\mathbf{x})$ (or concave for maximization), then the convergence guarantees applicable to smooth functions can still hold. In our case the function $U$ and $R$ for each agent can be the following:
\begin{subequations}
    \begin{equation}
        \begin{split}
             U_{i}(x) = & u_i(x) + \mathbbm{1} \{x \geq x_0\} \cdot  \\
             & \cdot \left( \frac{\partial u_i}{\partial x(i)} \bigg|_{x_0^{-}(i)} - \frac{\partial u_i}{\partial x(i)} \bigg|_{x_0^{+}(i)} \right) \cdot (x-x_0) 
        \end{split}
    \end{equation}
    \begin{equation}
        \begin{split}\
            R_{i}(x) = & - \mathbbm{1} \{x \geq x_0\} \cdot  \\ 
            & \cdot \left( \frac{\partial u_i}{\partial x(i)} \bigg|_{x_0^{-}(i)} - \frac{\partial u_i}{\partial x(i)} \bigg|_{x_0^{+}(i)} \right) \cdot (x-x_0) .
        \end{split}
    \end{equation}
\end{subequations}
In addition, the authors of \cite{SCA_Scutari_Theory_Part1} emphasize that the first construction rule, i.e., the strong convexity of the surrogate function, is the most critical. This condition significantly influences the choice of the optimization method used to solve the inner surrogate problem in \ref{eq:Surrogate_Opt_Problem}. The remaining construction rules serve as technical requirements to ensure that the surrogate function maintains the same local first-order behavior as the original function. However, due to the technical challenge introduced by the non-smoothness at the reference point in our problem, we introduce two additional construction conditions into our methodology to better handle this issue.
\begin{enumerate}[resume]
    \item $\Tilde{f}(\mathbf{x} | \mathbf{x}^{(k)}) \geq f(\mathbf{x}), \; \forall \; \mathbf{x} \in \mathcal{X}$ 
    \item If $x^{(k)}(i) \to x_0^(i)$, then $\frac{\partial \Tilde{u}_i}{\partial x(i)} \bigg|_{x_0^{+}(i)} \to \frac{\partial u_i}{\partial x(i)} \bigg|_{x_0^{+}(i)}$ and  $\frac{\partial \Tilde{u}_i}{\partial x(i)} \bigg|_{x_0^{-}(i)} \to \frac{\partial u_i}{\partial x(i)} \bigg|_{x_0^{-}(i)}$
\end{enumerate}

Regarding the convergence of the SCA algorithm, the authors of \cite{SCA_Scutari_Theory_Part1} prove that if all four surrogate function conditions are satisfied, and the step size is chosen to meet the following criteria:
\begin{equation}
     \theta^{(l)} \in (0,1], \theta^{(l)} \xrightarrow{l \to \infty} 0, \sum_{l} \theta^{(l)} = + \infty
\end{equation}
then every limit point of the sequence ${ \mathbf{x}^{(k)} }$ is a stationary point of the original optimization problem. In other cases, if the three basic conditions are satisfied, at least one feasible point of the sequence ${ \mathbf{x}^{(k)} }$ is a stationary point of the original optimization problem. In our case, the fourth condition does not hold due to the fact that the objective function is not smooth at the reference point.
\begin{algorithm}
\caption{SCA algorithm for optimization problem}\label{alg:SCA}
\begin{algorithmic}
    \STATE Set $l=0$, initialize feasible point $\mathbf{P}^{(0)} \in \mathcal{S}, \text{ step size }\{\theta^{(l)}\} \in (0,1]$
    \REPEAT
    \STATE $\hat{\mathbf{P}} \left( \mathbf{P}^{(l)} \right) =  \text{arg} \min  \Tilde{f} \left( \mathbf{P} | \mathbf{P}^{(l)} \right)$ \\
            \hspace{2cm} \text{s.t.} \hspace{0.1cm} {$\mathbf{P} \in \mathcal{S}$} \\
            \hspace{2.8cm}  $g \left(\mathbf{P}\right) \leq 0$ \\
    \STATE $\mathbf{P}^{(l+1)} = \mathbf{P}^{(l)} + \theta^{(l)} \cdot \left( \hat{\mathbf{P}} \left( \mathbf{P}^{(l)} \right) - \mathbf{P}^{(l)} \right)$
    \STATE $l \leftarrow l + 1$
    \UNTIL \text{convergence} \\
    \STATE \textbf{return} $\mathbf{P}^{(l)}$
\end{algorithmic}
\end{algorithm}
\begin{equation}
    \begin{aligned}
        \min_{\mathbf{P}} \quad & \Tilde{f} \left( \mathbf{P} | \mathbf{P}^{(l)} \right)\\
        \textrm{s.t.} \quad & \mathbf{P} \in \mathcal{S}\\
          &  g \left(\mathbf{P}\right) \leq 0  \\
    \end{aligned}
    \label{eq:Surrogate_Opt_Problem}
\end{equation}

We now proceed to apply the surrogate function construction rules within our proposed framework. The general form of the surrogate functions, tailored to the structure of our problem, is defined separately for the gain and loss subdomains.
For the gain subdomain, the surrogate utility function is given by:
\begin{equation}
    \Tilde{u}_{\textrm{gain}}(x(i)|x^{(l)}(i)) = \lambda_{\textrm{gain}} \cdot \frac{1 - \exp{ \left(\alpha_{\textrm{gain}} \cdot \frac{x(i) - \mu_{\textrm{gain}}}{m} \right)}}{ \alpha_{\textrm{gain}}} 
\end{equation}
and for the loss subdomain, the corresponding surrogate function is:
\begin{equation}
    \Tilde{u}_{\textrm{loss}}(x(i)|x^{(l)}(i)) = \lambda_{\textrm{loss}} \cdot \frac{1 - \exp{ \left(\beta_{\textrm{loss}} \cdot \frac{x(i) - \mu_{\textrm{loss}}}{n} \right)}}{ \beta_{\textrm{loss}}} 
\end{equation}

More precisely, we categorize all possible cases into six distinct scenarios, based on the values of the utility function parameters. \textbf{Case 1}: if the parameters satisfy $\frac{\alpha(i)}{\gamma_1(i)}<0,\frac{\beta(i)}{\gamma_2(i)}<0$, then the utility function is strictly concave over the entire domain. In this case, the original function itself serves as a valid surrogate, as it inherently satisfies all the construction rules. \textbf{Case 2}: if $\SNR_0(i) \leq \SNR^{(l)}(i)$ and $\frac{\alpha(i)}{\gamma_1(i)}<0,0 \leq \frac{\beta(i)}{\gamma_2(i)}$, the loss subdomain must be approximated using a concave surrogate function $\Tilde{u}_{\textrm{loss}}(x(i)|x^{(l)}(i))$ with $\mu_{\textrm{loss}} = x_0(i)$, $\beta_{\textrm{loss}} \leq 0$, and $\lambda_{\textrm{loss}} = - n \cdot \frac{\partial u_i}{\partial x(i)}\bigg|_{x_0^-(i)}$. The gain subdomain remains unchanged, since it is already concave.
An interesting observation is that the 'best' concave approximation of a convex function, given a specific intersection point and subject to the surrogate construction rules, is the tangent line at that point. This choice minimizes the Euclidean distance between the original and surrogate functions locally. However, it is important to emphasize that if a linear surrogate function (such as the tangent) is used, the first construction rule (strong convexity) is not satisfied. As a result, the PSM cannot be applied in the inner minimization step of SCA under this choice.

\textbf{Case 3}: if the parameters satisfy $\SNR_0(i) < \SNR^{(l)}(i)$ and $0 \leq \frac{\alpha(i)}{\gamma_1(i)}, \frac{\beta(i)}{\gamma_2(i)} \in \mathbb{R}$, 
both the gain and loss subdomains require surrogate approximations. For the gain subdomain, we define the surrogate function $\Tilde{u}_{\textrm{gain}}(x(i)|x^{(l)}(i))$ with $\mu_{\textrm{gain}} = x^{(l)}(i)$, $\alpha_{\textrm{gain}} \leq 0$, $l_{\textrm{gain}} = u_i(x^{(l)}(i))$, $\lambda_{\textrm{gain}} = - m \cdot \frac{\partial u_i}{\partial x(i)}\bigg|_{x^{(l)}(i)}$ and the loss subdomain with the following surrogate function $\Tilde{u}_{\textrm{loss}}(x(i)|x^{(l)}(i))$ with $\mu_{\textrm{loss}} = x_0(i)$ , $\beta_{\textrm{loss},1} \leq \min \left\{0, \frac{\beta(i)}{\gamma_2(i)} \right \}$, $|\beta_{\textrm{loss},2}| \leq |\beta(i)|$ and $0 < \beta_{\textrm{loss},2} \cdot \beta(i)$, $l_{\textrm{loss}} = \Tilde{u}_{\textrm{gain}}(x_0(i)|x^{(l)}(i))$, $\lambda_{\textrm{loss}} = - n \cdot \frac{\beta_{\textrm{loss},2}}{\beta_{\textrm{loss},1}} \cdot \max \left\{ \frac{\partial u_i}{\partial x(i)} \bigg|_{x^-_0(i)} , \frac{\partial \Tilde{u}_{\textrm{gain}}}{\partial x(i)} \bigg|_{x^+_0(i)} \right\}$.

\textbf{Case 4}: if $\SNR^{(l)}(i) \leq \SNR_0(i)$ and $0 \leq \frac{\alpha(i)}{\gamma_1(i)}, \frac{\beta(i)}{\gamma_2(i)} < 0$, we approximate the gain subdomain using a concave surrogate function $\Tilde{u}_{\textrm{gain}}(x(i)|x^{(l)}(i))$
with parameters $\mu_{\textrm{gain}} = x_0(i)$, $\alpha_{\textrm{gain}} \leq 0$ and $\lambda_{\textrm{gain}} = - m \cdot \frac{\partial u_i}{\partial x(i)}\bigg|_{x_0^+(i)}$. The loss subdomain is concave and therefore kept unchanged. \textbf{Case 5}: if $\SNR^{(l)}(i) < \SNR_0(i)$ and $\frac{\alpha(i)}{\gamma_1(i)} \in \mathbb{R}, 0 \leq \frac{\beta(i)}{\gamma_2(i)}$, both subdomains require surrogate approximations. The loss subdomain is approximated using the surrogate function $\Tilde{u}_{\textrm{loss}}(x(i)|x^{(l)}(i))$
with $\mu_{\textrm{loss}} = x^{(l)}(i)$ , $\beta_{\textrm{loss}} \leq 0$, $l_{\textrm{loss}} = u_i(x^{(l)}(i))$, $\lambda_{\textrm{gain}} = - n \cdot \frac{\partial u_i}{\partial x(i)}\bigg|_{x^{(l)}(i)}$
and the gain subdomain with the following surrogate function $\Tilde{u}_{\textrm{gain}}(x(i)|x^{(l)}(i))$
with $\mu_{\textrm{gain}} = x_0(i)$ , $\alpha_{\textrm{gain},1} \leq 0$ if $0 \leq\frac{\alpha(i)}{\gamma_1}$ or $\frac{\alpha(i)}{\gamma_1} \leq \alpha_{\textrm{gain},1} \leq 0$ if $\frac{\alpha(i)}{\gamma_1} < 0$, $|\alpha(i)| \leq |\alpha_{\textrm{gain},2}|$ and $0 < \alpha_{\textrm{gain},2} \cdot \alpha(i)$, $l_{\textrm{gain}} = \Tilde{u}_{\textrm{loss}}(x_0(i)|x^{(l)}(i))$, $\lambda_{\textrm{gain}} = - m \cdot \frac{\alpha_{\textrm{gain},2}}{\alpha_{\textrm{gain},1}} \cdot \min \left\{ \frac{\partial u_i}{\partial x(i)} \bigg|_{x^+_0(i)} , \frac{\partial \Tilde{u}_{\textrm{loss}}}{\partial x(i)} \bigg|_{x^-_0(i)} \right\}$.

\textbf{Case 6}: if $\SNR^{(l)}(i) = \SNR_0(i)$ and $0 \leq \frac{\alpha(i)}{\gamma_1(i)}, 0 \leq \frac{\beta(i)}{\gamma_2(i)}$, both subdomains must be approximated using concave surrogate functions centered at the reference point.
For the loss subdomain, we use the surrogate function $\Tilde{u}_{\textrm{loss}}(x(i)|x^{(l)}(i))$
with $\mu_{\textrm{loss}} = x_0(i)$ , $\beta_{\textrm{loss}} \leq 0$ and $\lambda_{\textrm{loss}} = - n \cdot \frac{\partial u_i}{\partial x(i)}\bigg|_{x_0^-(i)}$. For the gain subdomain, we define the surrogate function $\Tilde{u}_{\textrm{gain}}(x(i)|x^{(l)}(i))$
with $\mu_{\textrm{gain}} = x_0(i)$ , $\alpha_{\textrm{gain}} \leq 0$ and $\lambda_{\textrm{gain}} = - m \cdot \frac{\partial u_i}{\partial x(i)}\bigg|_{x_0^+(i)}$.
This configuration ensures that the surrogate utility function is piecewise concave, follows the same first-order behavior around the reference point as the original function, and satisfies the SCA construction rules required for convergence. As both subdomains start at the same point, this case also provides a numerically stable approximation around the reference threshold.

\subsection{Projected Subgradient Method}
Regarding the optimization problem \ref{eq:Surrogate_Opt_Problem}, we use the Lagrangian relaxation approach, following the methodology outlined in \cite{boyd-book}. To start with, we define the Lagrangian of the problem
\begin{equation}
    \mathcal{L}(\mathbf{P},k) = \Tilde{f} \left( \mathbf{P} | \mathbf{P}^{(l)} \right) + k \cdot g\left(\mathbf{P}\right) \;,\; 0 \leq k
\end{equation}
Additionally, the Lagrangian function admits a unique minimizer over $\mathbf{P}$, which is denoted $\mathbf{P}^*(k)$. The dual problem then consists in maximizing this dual function with respect to the dual variables:
\begin{equation}
    h(k) = \inf_{\mathbf{P} \in \mathcal{S}} \mathcal{L}(\mathbf{P},k) = \Tilde{f} \left( \mathbf{P}^*(k) | \mathbf{P}^{(l)} \right) + k \cdot g\left(\mathbf{P}^*(k)\right) \;,\; 0 \leq k.
\end{equation}
The dual problem then consists in maximizing this dual function with respect to the dual variables:
\begin{equation}
    \begin{aligned}
        \max_{k} \quad & h(k)\\
        \textrm{s.t.} \quad & 0 \leq k\\
    \end{aligned}
    \label{eq:Dual_Problem}
\end{equation}
Since Slater’s condition holds, strong duality applies, and the primal problem can be solved by first finding the optimal dual point $k^*$, then recovering the primal solution as $\mathbf{P}^* = \mathbf{P}^*(k^*)$.

We solve the dual problem using the projected subgradient method \cite{boyd-book},
\begin{equation}
    k^{(i+1)} = \left( k^{(i)} - \zeta_i \cdot q^{(i)} \right)_+ \;,\; q^{(i)} \in \partial(-h)(k^{(i)})
\end{equation}
where $q^{(i)}$ is a subgradient with respect to $k$ at the $i$-th iteration of the outer optimization problem with objective function $h(k)$, and $\zeta_i$ is the step size. The step sizes are chosen to be square summable but not summable, i.e.,
\begin{equation}
    0 \leq \zeta_i, \quad \sum^{+\infty}_{i = 1 }\zeta_i^2 < + \infty, \quad \sum^{+\infty}_{i = 1 }\zeta_i = + \infty
\end{equation}
in order to ensure convergence of the method to an optimal point. 
This choice of step size is motivated by the fact that our objective function, which is the surrogate used in the SCA framework, is Lipschitz continuous, as it is concave on the right of $x{(l)}(i)$ and concave but bounded on the left.
Moreover, due to the specific form of $h(k)$, the subdifferential of $-h$ at any point contains only a single element, which implies that $h$ is differentiable, and its gradient is given by:
\begin{equation}
    \partial(-h)(k) = - g\left(\mathbf{P}^*(k)\right),
\end{equation}
and the projected subgradient method for the dual problem has the form
\begin{subequations}
    \begin{equation} \label{eq:Dual_Problem_min}
        \mathbf{P}^{(i)} = \text{arg}\min_{\mathbf{P} \in \mathcal{S}} \left( \Tilde{f} \left( \mathbf{P} | \mathbf{P}^{(l)} \right) + k^{(i)} \cdot g(\mathbf{P}) \right)
    \end{equation}
    \begin{equation}
        k^{(i+1)} = \left( k^{(i)} + \zeta_i \cdot g\left( \mathbf{P}^{(i)} \right) \right). 
    \end{equation}
\end{subequations}

Since the set $S$ is convex and the optimization problem \ref{eq:Dual_Problem_min} is convex, we can once again apply the projected subgradient method to solve it as follows:
\begin{equation}
    \mathbf{P}^{(j+1)} = \Pi_{\mathcal{S}} \left( \mathbf{P}^{(j)} - \eta_j \cdot w^{(j)} \right) 
\end{equation}
where $\Pi$ is the Euclidean projection on $\mathcal{S}$
\begin{equation}
    \Pi_{\mathcal{S}}(x_0) = \min_{x} \Vert x - x_0 \Vert_2 \;:\; x \in \mathcal{S}
\end{equation}
and as $\mathcal{S} = \mathbb{R}^N_+$, the projection is equivalent to 
\begin{equation}
    \Pi_{\mathcal{S}}(x_0) = (x_0)_+.
\end{equation}
The parameter $\eta_j$ denotes the step of the method and is again chosen as \emph{square summable but not summable}. Furthermore, $w^{(j)}$ represents the subgradient with respect to $\mathbf{P}$ at the $j$-th iteration of the inner optimization problem, whose objective function is $f\left(\mathbf{P}\right) + k^{(i)} \cdot g\left(\mathbf{P}\right)$.

\section{Simulation Results}
\begin{figure}
    \centering
    \includegraphics[width=0.75\linewidth]{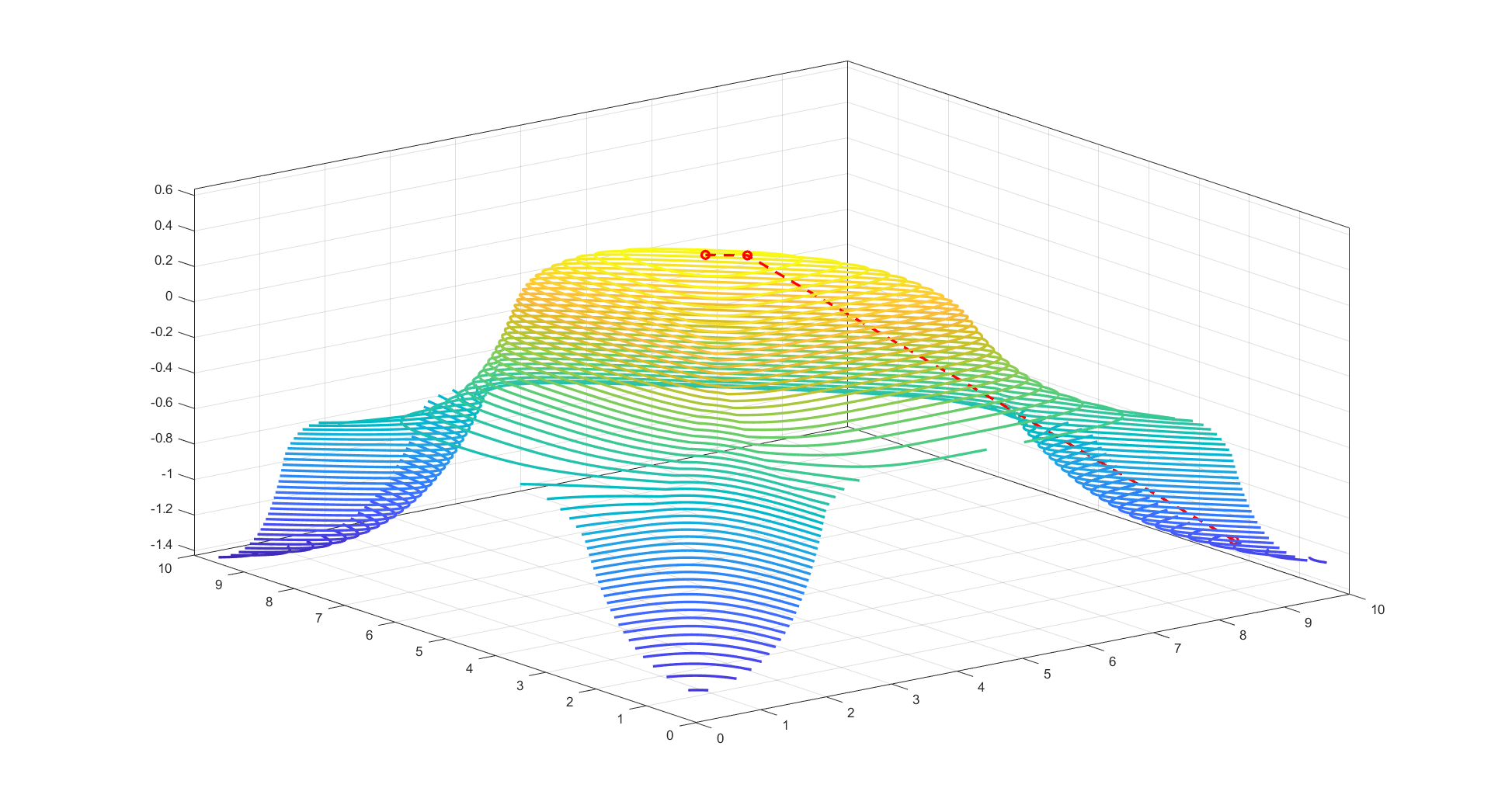}
    \caption{3D contour plot illustrating the convergence process for a scenario with $N=3$ agents.}
    \label{contour_convergence_1}
\end{figure}

We begin by solving the power allocation problem in a system with three agents experiencing Rayleigh fading, where risk behavior is modeled using an S-shaped CPT utility function. This simulation illustrates the convergence behavior of our proposed optimization framework through a simple yet representative scenario. As shown in the contour plot in Figure \ref{contour_convergence_1}, despite initializing near a local minimum, the algorithm quickly converges to a local maximum of the objective function.
Next, we simulate a system under three different scenarios of the agent population: $N = 10, 30, 50$. Each system is simulated over $500$ independent iterations. In each iteration, agents are assigned distinct CPT-based utility functions, and their channel gains are independently drawn from an exponential distribution with unit mean. The average SNR under equal power allocation is set to 7 dB.

To evaluate the performance of our proposed optimization method, we use as a metric the percentage of cases where our solution achieves an equal or better outcome, within a tolerance of $k$\%, compared to the solution obtained by MATLAB Sequential Quadratic Programming (SQP) Global Search toolbox. Specifically, a tolerance of $\pm k$\% implies that if the relative difference between our solution and MATLAB’s falls within this range, the two are considered equivalent. The results are summarized in Table \ref{tab:comparison_results}. We also report results based on the trimmed mean in Table \ref{tab:comparison_results_trimmean}.
\begin{table}
    \centering
    \begin{tabular} {|wc{0.3cm}|wc{2.5cm}|wc{2.5cm}|}
         \hline
         $N$ & \makecell{Mean \% better than \\ MATLAB (0\% tol.)} & \makecell{Mean \% better than \\ MATLAB (2\% tol.)}\\
         \hline
         10 & 38.9 & 49.8 \\
         \hline
         30 & 72.8 & 73.6 \\
         \hline
         50 & 96.2 & 96.4 \\
         \hline
    \end{tabular}
    \caption{Comparison of objective function values: proposed method vs. SQP (MATLAB).}
    \label{tab:comparison_results}
\end{table}

\begin{table}
    \centering
    \begin{tabular}{|m{0.3cm}|m{1.5cm}|m{1.5cm}|m{1.5cm}|m{1.5cm}|}
         \hline
         $N$ & Mean & 1\% Trimmed mean & 2\% Trimmed mean & 5\% Trimmed mean\\
         \hline
         10 & -10.3613 & -10.4444 & -10.5075 & -10.5354\\
         \hline
         30 & 40.6332 & 40.8660 & 41.2340 & 41.7988\\
         \hline
         50 & 81.3611 & 81.5843 & 81.6584 & 81.4217\\
         \hline
    \end{tabular}
    \caption{Trimmed mean of the percentage of quantitative better solution than SQP (MATLAB).}
    \label{tab:comparison_results_trimmean}
\end{table}
As shown in the results tables, our proposed method outperforms MATLAB's toolbox as the number of agents increases. However, this performance gain comes with a longer execution time. Specifically, our method is significantly slower for $N=10$ and $N=30$. Nevertheless, at $N = 50$, the gap narrows, with our approach being approximately four times slower. This trend suggests that as the agent population increases, our method not only maintains its performance advantage over MATLAB but also achieves comparable execution times.

\section*{Acknowledgment}
This work is partially supported by the European Research Council (ERC) under the EU’s Horizon 2020 research and innovation programme (Grant agreement No. 101003431), which partially supports the work of M. Kountouris. The work of S. Vaidanis and P. A. Stavrou is supported by the SNS JU project 6G-GOALS \cite{strinati:2024} under the EU’s Horizon programme Grant Agreement No. 101139232. S. Vaidanis is also supported by the Onassis Foundation (Scholarship ID: F ZU 076-1/2024-2025).  

\bibliographystyle{IEEEtran}
\bibliography{my_bibliography}

\end{document}